\documentclass[10pt,twocolumn,hidelinks]{article}
\usepackage[letterpaper, total={7.0in, 9.5in}]{geometry}

\usepackage[doi, etalmode=truncate, maxauthors=30]{achemso}
\setkeys{acs}{usetitle=true} 

\usepackage{graphicx} 
\usepackage[version=4]{mhchem}
\usepackage{amsmath,amsfonts,amssymb}
\usepackage{physics}
\usepackage{xfrac}
\usepackage{bm}
\usepackage{array}
\usepackage{titlesec}
\usepackage{balance}
\usepackage[dvipsnames]{xcolor}  
\usepackage{xspace}  
\usepackage{units}
\usepackage{url}
\usepackage[hidelinks]{hyperref}
\usepackage{enumitem}
\usepackage{verbatim}
\usepackage{booktabs}

\usepackage{setspace}
\usepackage[font=small,labelfont=bf]{caption}

\titlespacing{\section}{0pt}{3ex}{1ex}
\titlespacing{\subsection}{0pt}{2ex}{1ex}
\titlespacing{\subsubsection}{0pt}{2ex}{1ex}

\begin{document}

\renewcommand{\thesection}{\arabic{section}}
\renewcommand{\thesubsection}{\arabic{subsection}}
\renewcommand{\thesubsubsection}{\arabic{subsubsection}}
\titleformat{\section}
  {\sffamily\bfseries\large}{\thesection.}{0.5em}{}
\titleformat{\subsection}
  {\sffamily\bfseries}{\thesection.\thesubsection}{0.5em}{}
\titleformat{\subsubsection}
  {\sffamily}{\thesection.\thesubsection.\thesubsubsection}{0.5em}{}




\def\blue{\textcolor{blue}}
\def\red{\textcolor{red}}
\def\green{\textcolor{Green}}
\def\orange{\textcolor{orange}}

\def\degC{$^\circ$C\xspace}
\def\Mw{$M_\text{w}$\xspace}
\def\Mn{$M_\text{n}$\xspace}
\def\MwMn{$M_\text{w}/M_\text{n}$\xspace}
\def\Tg{$T_\text{g}$\xspace}
\def\Tgbulk{$T_\text{g}^\text{bulk}$\xspace}
\def\Tgh{$T_\text{g}(h)$\xspace}
\def\Tgz{$T_\text{g}(z)$\xspace}
\def\TgZero{$T_\text{g}(z=0)$\xspace}

\twocolumn[   

\renewcommand{\baselinestretch}{1.5}
\sffamily
\vspace*{2mm}

\noindent \textbf{\Large End-Tethered Chains Increase the Local Glass Transition Temperature of \\ Matrix Chains by 45 K Next to Solid Substrates Independent of Chain Length}

\vspace{3mm}

\renewcommand{\baselinestretch}{1.0} \normalsize

\noindent \textbf{James H.\ Merrill, Ruoyu Li, and Connie B.\ Roth*}

\noindent Department of Physics, Emory University, Atlanta, Georgia 30322 USA

\vspace{3mm}

\noindent Dated: November 23, 2022,
\hspace{5mm} $^*$Email contact: cbroth@emory.edu


\vspace{5mm}
\rmfamily \normalsize


\noindent \textbf{\sffamily Abstract: }  
The local glass transition temperature \Tg of pyrene-labeled polystyrene (PS) chains intermixed with end-tethered PS chains grafted to a neutral silica substrate was measured by fluorescence spectroscopy.  To isolate the impact of the grafted chains, the films were capped with bulk neat PS layers eliminating competing effects of the free surface.  Results demonstrate that end-grafted chains strongly increase the local \Tg of matrix chains by $\approx$45~K relative to bulk \Tg, independent of grafted chain molecular weight from \Mn = 8.6 to 212 kg/mol and chemical end-group, over a wide range of grafting densities $\sigma = 0.003$ to 0.33 chains/nm$^2$ spanning the mushroom-to-brush transition regime.  The tens-of-degree increase in local \Tg resulting from immobilization of the chain ends by covalent bonding in this athermal system suggests a mechanism that substantially increases the local activation energy required for cooperative rearrangements.

\vspace*{5mm}
]  

\newcommand{\quickwordcount}[1]{%
  \immediate\write18{texcount -1 -q #1.tex > #1-words.sum }%
  \input{#1-words.sum} words }


Grafted chains are widely used to tune the adhesion and lubrication of interfaces and alter the physical properties of polymer matrices.\cite{LegerCreton2008, FettersScience1994, BrochardWyartMacro1996, CasoliReiterLangmuir2001}  Polymer nanocomposites (PNCs) frequently use grafted chains to improve matrix adhesion and dispersion of nanoparticles.\cite{KumarWineyMacro2017, KumarMacro2013, GanesanJayaramanSM2014, MaillardNanoLett2012, LakkasSoftMatter2021}  However, the precise mechanism by which material properties of polymer matrices are altered in the vicinity of grafted chains is not well-understood, especially those associated with the glass transition.\cite{KumarMacro2013, KumarJCP2017review, BansalJPCB2006, AkcoraMacro2010, BerriotSottaMacro2002, OhGreenNatMater2009, PaponPRL2012, HoltSokolovACSNano2016, SakibSimonMacro2020, AskarTorkelsonMacro2017, WeiTorkelsonMacro2020}  Part of the difficulty is that there are limited experimental techniques capable of probing local material properties that can provide direct insight into how matrix chains next to the tailored interface are changed.  In PNCs, the impact grafted chains at a solid surface have are characterized by several interdependent quantities related to nanoparticle geometry and preparation methods:  grafting density, grafted and matrix chain lengths, as well as their degree of interpenetration, surface curvature of the nanoparticle, not to mention their loading fraction and spatial dispersion.  The geometrical complications of PNCs can be simplified by making comparisons to thin films with similar interfacial interactions.\cite{BansalNatMater2005, RittigsteinNatMater2007, KropkaGanesanPRL2008, StarrJCP2019}  In this vein, we use a planar film geometry with a localized fluorescence method to characterize the local glass transition temperature of matrix chains next to the substrate interface \TgZero as a function of grafting density and chain length of end-grafted chains, providing insight into the underlying mechanism responsible for the large observed local \Tg increase.  

Attempts to characterise the impact of grafted chains on the film-average \Tgh as a function of decreasing film thickness $h$ in polystyrene (PS) films date back to Keddie et al.\ in 1995.\cite{KeddieIsraelJChem1995}  An in-depth analysis of the existing literature by Restagno and coworkers in 2017,\cite{HenotEPJE2017} concluded that grafted chains appear to have little to no impact on \Tgh.\cite{TateJCP2001, CloughMacro2011, GanesanMacro2010, ZuoSoftMatter2017, GanesanMacro2018}  However, measurements by Lan and Torkelson showed that the local \Tg of end-grafted polymer brushes could have a gradient of 50~K across the depth of the film, being strongly reduced near the free surface and increased near the substrate interface.\cite{LanTorkelsonPolym2015}  This would suggest that the strong free surface effect can mask a large \Tg increase near the substrate interface due to grafted chains.  

Huang and Roth avoided the competing effects of the free surface, using localized fluorescence to measure the local glass transition temperature \Tgz within a PS matrix as a function of distance $z$ from a silica substrate with end-tethered PS chains,\cite{HuangACSMacroLett2018} where the bare silica substrates ($\sigma = 0$) do not impart any local \Tg perturbation.\cite{EllisonNatMater2003,HuangJCP2020}  These results demonstrated a large \TgZero increase as high as $49 \pm 2$~K above bulk \Tg of the PS matrix (\Tgbulk = 100 \degC) for end-tethered monocarboxy-terminated polystyrene (PS-COOH) with a molecular weight of \Mn = 98.8 kg/mol (\MwMn = 1.03) at a grafting density $\sigma = 0.011$ chains/nm$^2$.  They identified that the highest \TgZero increase occurred at an optimum grafting density in the middle of the mushroom-to-brush transition region when the matrix chains still had good ability to interpenetrate with the end-grafted chains.\cite{HuangACSMacroLett2018}  In the strong brush limit at very high grafting densities ($\sigma \rightarrow \infty$), the matrix chains would be expected to recover \Tgbulk because they would be incapable of penetrating into the tight brush, resulting in an effectively neutral interface.  A similar nonmonotonic change in \Tgh with grafting density was also observed by Lee et al.\cite{GanesanMacro2018} 

\begin{figure*}[tb!]  
    \centering
    \includegraphics[width=7.0in]{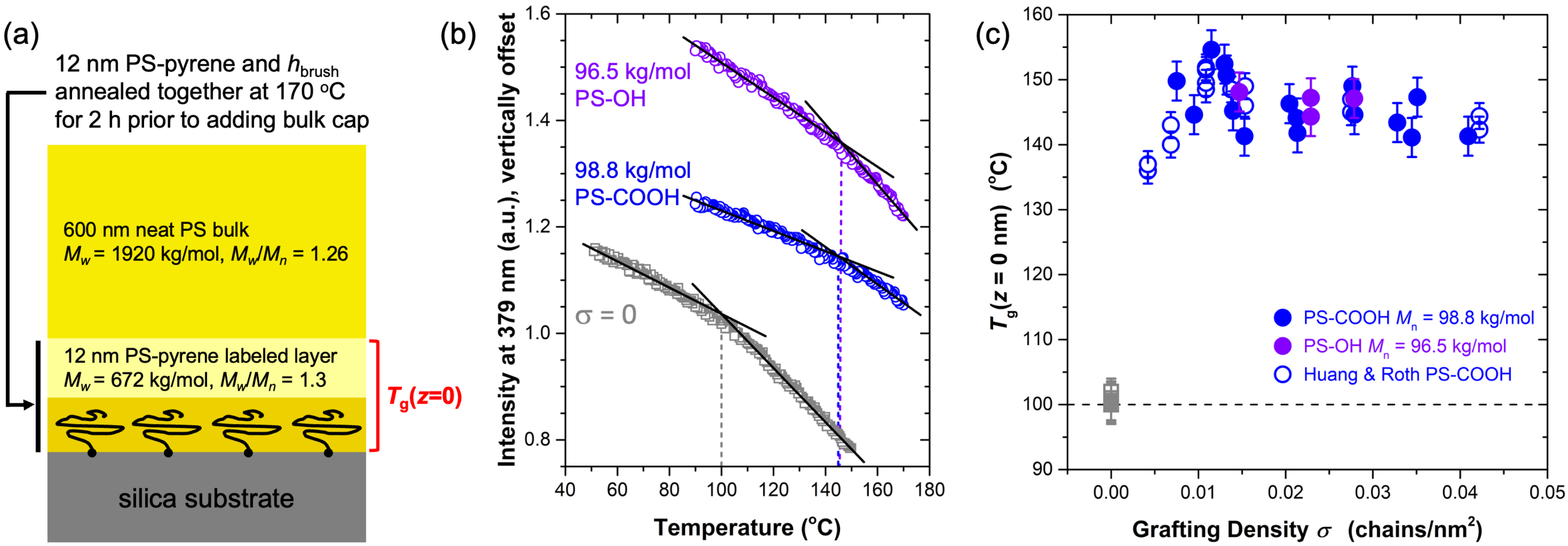}
    \caption{(a) Multilayer geometry assembled to measure the local glass transition temperature \TgZero next to end-grafted silica substrates where a 12-nm thick pyrene-labeled PS probe layer is intermixed with the end-grafted chains, prior to capping the sample with a bulk neat PS layer to isolate the fluorescent layer from the free surface.  (b) Temperature dependence of normalized fluorescence intensity of pyrene-labeled PS probe layers intermixed with end-grafted chains of 100k PS-COOH ($\sigma = 0.022$ chains/nm$^2$, blue circles), 100k PS-OH ($\sigma = 0.021$ chains/nm$^2$, purple circles), and next to a bare silica substrate ($\sigma = 0$, gray squares), data sets are vertically offset for clarity.  (c) Local \TgZero as a function of grafting density $\sigma$ next to end-grafted chains of 100k PS-COOH from both this study (filled blue circles) and Ref.~\citenum{HuangACSMacroLett2018} (open blue circles), and 100k PS-OH (filled purple circles), relative to bare silica substrates (gray squares from both this study and Ref.~\citenum{HuangACSMacroLett2018} at $\sigma = 0$). }
    \label{Fig1}
\end{figure*}

As the grafting densities $\sigma$ over which the mushroom-to-brush transition occurs depend on the length $N$ of the end-grafted chains, one might expect a molecular weight dependence to how \TgZero varies with grafting density.  Often characterized by the reduced tethered density $\Sigma = \pi R_\text{g}^2 \, \sigma$, the mushroom-to-brush transition region typically occurs in the range $1 < \Sigma \lesssim 5$--10,\cite{BrittainJPolymSciA2007,Cheng2008book} where the radius of gyration $R_\text{g} \sim N^{\sfrac{1}{2}}$ as the end-tethered chains are still in their ideal conformations at these low grafting densities.\cite{JonesRichardsBook, AubouyMacro1995}   In the present work, we investigate how the grafting density dependence of \TgZero varies with the molecular weight of the end-grafted chains, varying \Mn from 8.6 kg/mol to 212 kg/mol (as specified in \textbf{Table~\ref{Table1}}).    

\begin{table}[b!]  
\caption{Number average molecular weight \Mn and dispersity \MwMn of the end-grafted polystyrene investigated, terminated with either a carboxy (\ce{-COOH}) or hydroxy (\ce{-OH}) end-group, along with their designation.}  \label{Table1}
\centering
\begin{tabular}{lrr} \toprule
polymer designation & \Mn (g/mol) & \MwMn \\ \midrule
9k PS-OH & 8,600 & 1.08 \\
14k PS-COOH & 13,600 & 1.07 \\
50k PS-COOH & 45,700 & 1.07\\ 
100k PS-OH & 96,500  & 1.12 \\
100k PS-COOH & 98,800 & 1.03 \\
200k PS-COOH & 212,000 & 1.08 \\
\bottomrule
\end{tabular}
\end{table}

The multilayer geometry depicted in \textbf{Figure~\ref{Fig1}a} illustrates how we measured the local glass transition temperature \TgZero of PS matrix chains next to PS end-grafted silica substrates.  End-grafted substrates were made by annealing films of either PS-COOH or PS-OH on piranha cleaned silica substrates at 170 \degC for 30 min to cause chemical grafting, and then bathed in 90 \degC toluene for 20 min to wash away ungrafted chains.\cite{HuangACSMacroLett2018, CloughMacro2011, KeddieIsraelJChem1995, TateJCP2001}  Grafting density was varied by reducing the initial thickness of the PS-COOH or PS-OH films.  After washing, the final dry brush thickness $h_\text{brush}$ was determined by ellipsometry to determine the grafting density $\sigma = \frac{\rho N_A h_\text{brush}}{M_\text{n}}$, where $\rho = 1.045$ g/cm$^3$ was taken as the bulk density of PS,\cite{Orwoll2007Markhandbook} $N_A$ is Avogadro's number, and \Mn is the number average molecular weight of the end-grafted chains.  Fluorescent probe layers of pyrene-labeled PS (\Mw = 672 kg/mol, \MwMn = 1.3, 1.4~mol\% pyrene\cite{HuangACSMacroLett2018, BaglayJCP2015, BaglayJCP2017}) with 12-nm thickness were initially spin-coated onto mica and then floated atop the end-grafted substrates.  These two layers were then annealed together at 170 \degC for 2~h to ensure good interpenetration of the pyrene-labeled matrix chains with the end-tethered chains.\cite{HuangACSMacroLett2018, ClarkePolym1996, ChenneviereMacro2013, OConnorMcLeishMacro1993}  A neat PS (\Mw = 1920 kg/mol, \MwMn = 1.26) layer of bulk thickness ($\approx$600~nm) was then floated on top to avoid any impact of the free surface on the fluorescent probe layer.  Immediately prior to the fluorescence measurements the samples were further annealed at 130 \degC for 20 min to consolidate the multilayer stack into a single material with no air gaps.  This step was also used to remove thermal history and equilibrate the samples prior to measuring \Tg, where the fluorescence intensity of pyrene was monitored at the first emission peak ($\lambda = 379$~nm) on cooling at 1 K/min from 170 \degC.\cite{RauscherMacro2013, BaglayJCP2015, HuangACSMacroLett2018}  

Pyrene fluorescence is well-established as a method of measuring \Tg based on its sensitivity to the density, polarity, and stiffness of the dye molecule's local environment.\cite{EllisonNatMater2003, RauscherMacro2013}  The resulting temperature dependence reflects the relative probability of radiative (fluorescent) versus nonradiative decay, exhibiting a transition as the local environment vitrifies, where the intersection of linear fits to the temperature-dependent intensity in the liquid and glassy states corresponds to \Tg.  \textbf{Figure~\ref{Fig1}b} shows representative data sets of the temperature-dependent fluorescence intensity for 100k PS-COOH (\Mn = 98.8 kg/mol) at a grafting density of $\sigma = 0.022$ chains/nm$^2$ and 100k PS-OH (\Mn = 96.5 kg/mol) at $\sigma = 0.021$ chains/nm$^2$ demonstrating the large increase in local \Tg that both these end-tethered chains produce:  \TgZero = $145 \pm 3$ \degC for PS-COOH and \TgZero = $146 \pm 3$ \degC for PS-OH.  This is notably in strong contrast to the \TgZero = $100 \pm 2$ \degC observed for the bare silica substrate ($\sigma = 0$), which is consistent with \Tgbulk for PS.  The grafting density dependence of \TgZero is plotted in \textbf{Figure~\ref{Fig1}c}, where we demonstrate good reproducibility between the present study's measurements of 100k PS-COOH with those from Ref.~\citenum{HuangACSMacroLett2018}.  The previously observed\cite{HuangACSMacroLett2018} peak in \TgZero occurs in the same location, at $\sigma = 0.011$ chains/nm$^2$, but appears less distinctive now with the additional data.  Thus, we conclude that all grafting densities measured with 100 kg/mol grafted chains from $\sigma = 0.004$ to $0.042$ chains/nm$^2$, which broadly span the mushroom-to-brush transition regime, result in primarily the same strongly elevated local \Tg of the matrix chains next to the substrate interface.  

\begin{figure}[t!]  
    \centering
    \includegraphics[width=3.3in]{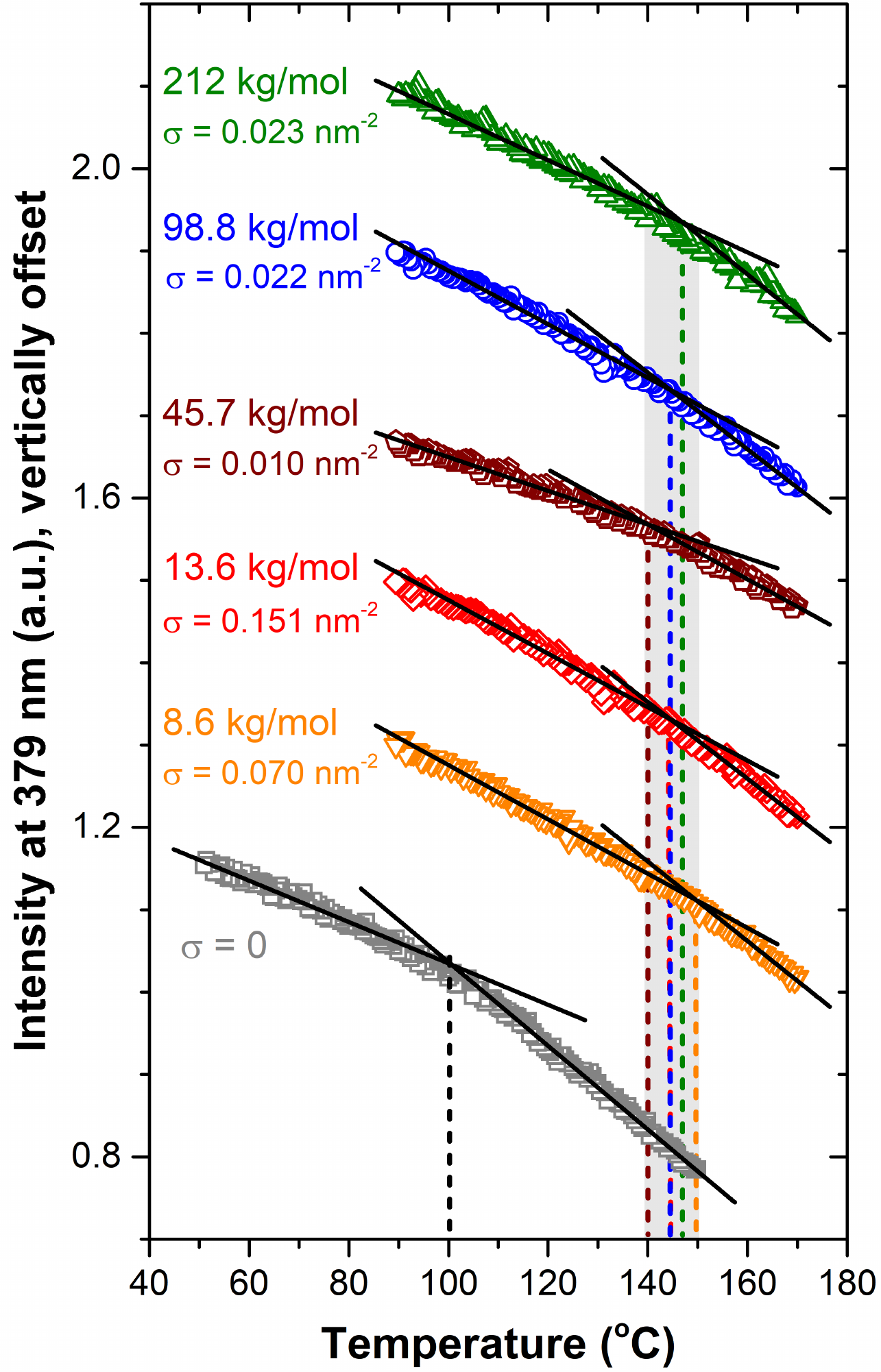}
    \caption{Fluorescence intensity vs.\ temperature measured for pyrene-labeled PS chains intermixed with the end-grafted PS chains at the substrate interface demonstrating elevated \TgZero values of $145 \pm 4$ \degC for all end-grafted substrates.  The \Mn and $\sigma$ of the grafted chains are labeled for each data set, which have been normalized to their mean intensity and vertically shifted for clarity.}
    \label{Fig2}
\end{figure}

In \textbf{Figure~\ref{Fig2}}, we compare the impact of varying the length of the end-grafted chains on the local \TgZero increase, varying the molecular weight of the PS-COOH or PS-OH from \Mn = 8.6 kg/mol to 212 kg/mol as denoted in Table~\ref{Table1}.  Representative traces of the temperature dependent fluorescence intensity measured for the pyrene-labeled PS chains intermixed with the end-grafted PS chains next to the silica substrate interface are plotted for select grafting densities $\sigma$.  To within experimental error, we observed similarly elevated \TgZero values ($145 \pm 4$ \degC) for all the end-grafted chains shown in Fig.~\ref{Fig2} regardless of chain length.  For example, the 200k PS-COOH with \Mn = 212 kg/mol at $\sigma = 0.023$ chains/nm$^2$ results in a \TgZero = 148 $\pm$ 3 \degC, experimentally equivalent to the 100k PS-COOH with \Mn = 98.8 kg/mol at $\sigma = 0.022$ chains/nm$^2$ where \TgZero = 145 $\pm$ 3 \degC, despite the end-grafted chains being twice as long.  More remarkable is that even the very short end-grafted chains impart similarly strongly elevated local \Tg values to the intermixed pyrene-labeled PS matrix chains regardless if the grafting density is high or low:  14k PS-COOH with \Mn = 13.6 kg/mol at $\sigma = 0.151$ chains/nm$^2$ gives \TgZero = 145 $\pm$ 3 \degC and 9k PS-OH with \Mn = 8.6 kg/mol at $\sigma = 0.070$ chains/nm$^2$ gives \TgZero = 149 $\pm$ 3 \degC.  

\begin{figure}[tb!]  
    \centering
    \includegraphics[width=3.3in]{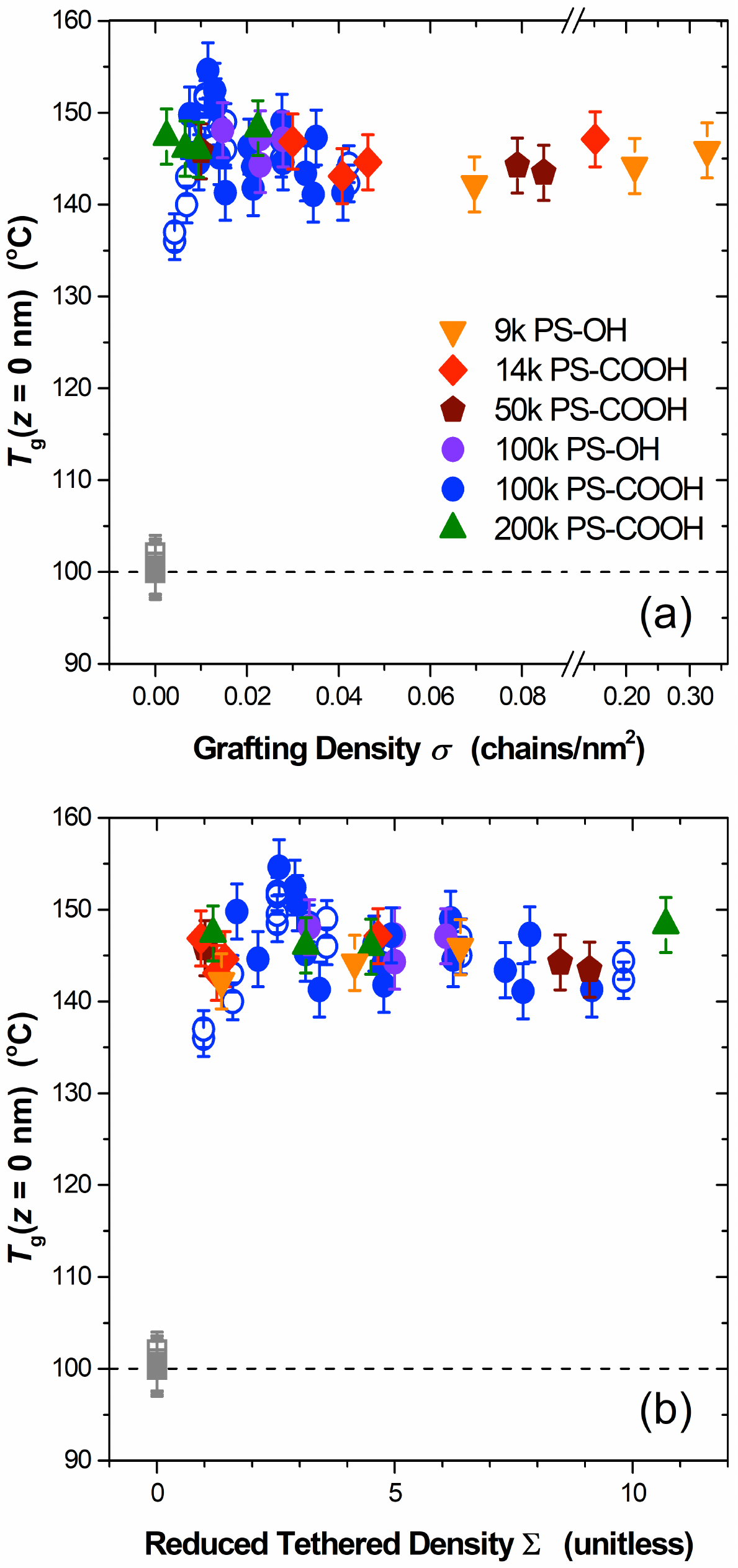}
    \caption{Local \TgZero of the pyrene-labeled PS matrix chains intermixed with the end-grafted PS-COOH or PS-OH chains next to the silica substrate interface plotted as a function of (a) grafting density $\sigma$ and (b) reduced tethered density $\Sigma = \pi R_\text{g}^2 \, \sigma$ that accounts for the greater surface coverage provided by the higher molecular weight grafted chains.  No obvious trend is observed with varying grafting density or molecular weight of the end-grafted chains, all imparting a consistently elevated \TgZero of 146 $\pm$ 3 \degC on average, relative to silica substrates with no grafted chains (gray squares). }
    \label{Fig3}
\end{figure}

\textbf{Figure~\ref{Fig3}a} plots all the measured \TgZero values as a function of grafting density $\sigma$ for the different molecular weights of end-grafted chains as indicated by symbol.  For all the grafting densities we investigated from $\sigma = 0.003$ to 0.33 chains/nm$^2$, the \TgZero values are strongly elevated ($146 \pm 3$ \degC) relative to the zero grafting density case of the bare silica substrate that reports a \TgZero = 100 $\pm$ 2 \degC equivalent to \Tgbulk.  The lack of grafting density dependence in \TgZero is surprising given that the surface coverage of the substrate interface by the end-grafted chains changes markedly with the molecular weight of end-grafted chains.  At a constant grafting density $\sigma$, the areal coverage of the surface increases linearly with the length $N$ of the end-grafted chains.  This difference in surface coverage can be accounted for by plotting the \TgZero values as a function of the reduced tethered density $\Sigma = \pi R_\text{g}^2 \, \sigma$ as shown in \textbf{Figure~\ref{Fig3}b}, where the grafting density $\sigma$ is multiplied by the projected area of the tethered chain onto the surface assuming a spherical shape with radius equal to the radius of gyration $R_\text{g}$.\cite{BrittainJPolymSciA2007, Cheng2008book}  For the low grafting densities we are investigating, the end-grafted chains are expected to have close to an ideal conformation when intermixed with the matrix chains.\cite{JonesRichardsBook, AubouyMacro1995}  Thus, $R_\text{g}$ values were calculated assuming ideal chain statistics as $R_g^2 = \frac{Nb^2}{6}$ with $b=0.67$~nm for PS.\cite{LodgeBook, FettersMacro1994}  

We stress that this strong increase in local \TgZero only occurs when PS-COOH or PS-OH chains are grafted to the silica substrate.  Several control measurements, described at length in Supporting Information, were performed to verify this point.  These include taking 100k neat PS, without the -COOH or -OH grafting groups, through all the same sample preparation steps resulting in noncovalent adsorbed layers, with the local \TgZero measured next to such adsorbed layers found to be equivalent to \Tgbulk.  These tests allow us to conclude that the strong increase in local \TgZero is the result of the covalent bond formed between the substrate and end-grafted PS-COOH or PS-OH chains, and not the result of some noncovalent adsorption or other sample processing condition.  

From the observed \TgZero results shown in Fig.~\ref{Fig3}, we can draw several conclusions about the possible underlying mechanism for the large increase in local \Tg caused by the end-grafted chains.  The observation that \TgZero does not vary with grafting density or grafted chain length indicates that the mechanism for the large increase in local \Tg of the matrix chains is not caused by a change in surface energy of the substrate resulting from either an increase in surface coverage of PS monomers or the grafted chain-end chemistry.  Similarly, \TgZero would also be expected to increase with increasing grafting density if the local \Tg increase were associated with a loss of free volume at the grafting site.\cite{TsuiWangJCP2020}  Previous efforts by our lab have not found local \Tgz changes in thin films to be correlated with density.\cite{HanJCP2021, HanJCP2020}  The magnitude of the large $\approx$45~K increase in local \Tg caused by the grafted chain-ends is also informative.  In their study of films with grafted chains, Lee et al.\ attributed the \Tgh increase to enhanced friction imparted by the brush as complementary computer simulations observed a factor of two slow down in segmental dynamics of matrix chains intermixed with end-grafted chains.\cite{GanesanMacro2018}  However, an increased segmental friction coefficient would not be expected to increase \Tg by more than a few degrees, as it would only modify the time scale for barrier hopping attempts.\cite{TalknerRevModPhys1990, RothChp1:2016}  

Immobilizing chain ends to the substrate interface in this athermal system is reminiscent of computer simulations that have demonstrated two-orders of magnitude increases in the local alpha-relaxation time $\tau_\alpha(T)$ next to ``rough'' walls created by immobilizing Lennard-Jones (LJ) particles on a surface,\cite{BaschnagelJPCM2005, ScheidlerEPL2022, ScheidlerJPCB2004, HanakataNatComm2014, HanakataJCP2015, LyulinJCP2015, SmithPRL2003} or studies that have investigated the impact of randomly pinned particles on cooperative dynamics.\cite{BerthierKobPRE2012, PhanJCP2018randpin, CammarotaPNAS2012}  For the range of grafting densities we investigated, $\sigma = 0.003$--0.33 chains/nm$^2$, the lateral spacing between grafting sites $x_\sigma = \sigma^{-\sfrac{1}{2}}$ is between 1.8 nm and 18 nm, roughly the same order of magnitude as the $\approx$3~nm characteristic length scale associated with cooperative motion in PS.\cite{HempelJPCB2000}  However, the chain connectivity of grafting to the substrate interface appears to create a larger effect than the pinning of disconnected LJ spheres.  Rather, the magnitude of the \Tg increase caused by grafting is comparable to that observed in associating polymers such as ionomers, vitrimers, and telechelic polymers where chains become tethered to cluster centers.\cite{RuanSimmonsMacro2015, GhoshSchweizerMacro2020, SingSchweizerMacro2022, GhoshMacro2022}  Recent theoretical efforts by Ghosh and Schweizer modeling associating polymers treated the ``sticky'' groups as causing persistent caging units, finding that the energy barrier for hopping could double for only 10\% sticky groups, leading to slower segmental dynamics by $\sim$9 orders of magnitude, equivalent to a \Tg increase of $\sim$30~K.\cite{GhoshSchweizerMacro2020}  How exactly chain connectivity alters the energy barrier for cooperative rearrangements is still an area of active research.\cite{RuanSimmonsMacro2015, GhoshSchweizerMacro2020, SokolovMacro2004, SchweizerMacromol2015, OlmstedPRX2022}

Given all these considerations, we conclude that chain connectivity to the substrate interface by covalent bonding is key to causing the large increase in local \Tg, yet we note that the pyrene dye is only attached to the intermixed homopolymer chains, not to the tethered chains themselves.  The large $\approx$45~K magnitude in local \Tg increase implies slower segmental dynamics by at least $\sim$10 orders of magnitude suggesting that grafting causes a substantial (at least factor of two) increase in the energy barrier associated with local cooperative rearrangements.  We are also left with several open questions.  Most significantly is the range of the \Tgz increase away from the grafted interface.  Huang and Roth observed that the \Tgz profile extended to $z \approx 100$--125 nm before \Tgbulk was recovered for the 100k PS-COOH tethered chains.\cite{HuangACSMacroLett2018}  Current theoretical understanding for how far interface perturbations to dynamics propagate would expect a much shorter distance.\cite{RuanSimmonsMacro2015, HanakataJCP2015, SchweizerSimmonsJCP2019, PhanACSMacroLett2020}  Future work will map how the extent of the \Tgz profile might vary with tethered chain length, which should shed light on this mechanism.  Also puzzling is the related work of Paeng and coworkers that used rotational fluorescence correlation microscopy of dyes tethered to $\approx$10 kg/mol chains, where no change in the $\tau_\alpha(T)$ dynamics were observed when these probe chains were grafted to the substrate interface.\cite{PaengMacro2021, PaengJCIS2023}  We speculate that this may be due to a discrepancy in how dynamic versus thermodynamic measures report \Tg near interfaces,\cite{AlcoutlabiJPCM2005, PriestleyJNCS2015, SimmonsJCP2017, StarrPNAS2018, HanSoftMatt2022} or from some difference in how disparate fluorescent dyes are sensitive to the altered local dynamics.\cite{PaengACSMacroLett2019, PaengMacro2022, EllisonNatMater2003, KimEPJE2009}  Future efforts will also work on investigating these possibilities.


\subsection*{Associated Content}
\noindent
\textbf{Supporting Information} available:  Details of experimental procedures and control measurements (PDF).  

\subsection*{Conflict of interest}
\noindent
The authors declare no competing financial interest.  

\subsection*{Acknowledgments}
\noindent
The authors gratefully acknowledge support from the National Science Foundation Polymers Program (DMR-1905782) and Emory University, as well as helpful discussions with Daniel Sussman. 
\balance
\renewcommand\refname{\small REFERENCES}
\footnotesize
\bibliography{references}

\clearpage

\newgeometry{margin=1in}
\setlength{\textwidth}{6.5in}       
\setlength{\textheight}{9in}        

\normalsize
\onecolumn
\doublespace
\fontsize{11}{12}\selectfont
\setlength{\parindent}{1cm}

\pagenumbering{arabic}      
\renewcommand*{\thepage}{S\arabic{page}}

\renewcommand\thefigure{S\arabic{figure}}    
\setcounter{figure}{0}  

\renewcommand\theequation{S\arabic{equation}}  
\setcounter{equation}{0}

\sffamily

\noindent SUPPORTING INFORMATION
\vspace{1mm}

\noindent
\textbf{\Large End-Tethered Chains Increase the Local Glass Transition Temperature of Matrix Chains by 45 K Next to Solid Substrates Independent of Chain Length}

\noindent
James H. Merrill, Ruoyu Li, and Connie B.\ Roth*

\noindent
\textit{Department of Physics, Emory University, Atlanta, Georgia, 30322 USA}

*To whom correspondence should be addressed.  Email: cbroth@emory.edu
\rmfamily 

\vspace{-3mm}
\section*{EXPERIMENTAL METHODS}

\subsection*{Sample preparation}

Grafting of monocarboxy-terminated polystyrene (PS-COOH) or monohydroxy-terminated polystyrene (PS-OH) chains on 25 mm $\times$ 25 mm substrates of both optical quality quartz for fluorescence and silicon for ellipsometry measurements were done in parallel together to ensure that both underwent the same grafting process.  This allowed us to determine the dry brush thickness for both substrates by measuring $h_\text{brush}$ with ellipsometry on the silicon substrates that provide better optical contrast and hence greater accuracy.  Immediately prior to spin-coating PS-COOH or PS-OH, the silicon and quartz substrates were immersed in a 1:2 mixture of 13 M hydrochloric acid (\ce{HCl}) and deionized (DI) water for at least 30 s, followed by a rinse in fresh DI water for 30 s, and then blown dry with \ce{N2} gas.  To vary the grafting density, PS-COOH or PS-OH films of varying thickness from sub-nanometer up to bulk were spin-coated and then annealed at 170 \degC for 30 min.  In our previous study grafting PS-COOH chains, 90 min at 170 \degC was used for the grafting time in accordance with existing literature.\cite{HuangACSMacroLett2018}  In our experimental tests, we found 30 min at 170 \degC to be sufficient, producing an equivalent dry brush thickness $h_\text{brush}$ to any longer grafting time up to 72 h.  After the grafting step at 170 \degC, the PS-COOH or PS-OH films were immersed in heated toluene at 90~\degC for 20 min, to wash away any ungrafted chains, then rinsed with acetone and DI water, while being blown dry with nitrogen gas after each step, as outlined by Huang and Roth.\cite{HuangACSMacroLett2018}.  Finally,  the washed, grafted substrates were dried under vacuum at room temperature overnight.  

To clean the optical quartz substrates between samples and ensure that any previously grafted or adsorbed chains were completely removed, the quartz substrates were piranha cleaned by immersing $\sim$15 of them at a time into a 3:1 mixture of concentrated \ce{H2SO4} heated to 60 \degC and 30 vol\% \ce{H2O2} for 20 min [1].  Substrates were then rinsed thoroughly with DI water, and stored in a jar of DI water until use.  We verified this procedure removed grafted chains by creating grafted substrates on silicon and demonstrating that zero thickness ($\lesssim$0.2 nm within error of the ellipsometer measurement [1]) was recovered after piranha cleaning. 

Ellipsometry (Woollam M-2000) was used to measure the dry brush thicknesses $h_\text{brush}$ of the resulting grafted layers on silicon produced simultaneously in the same batch with the quartz substrates for fluorescence.  Because the typical thicknesses of the grafted layers are below 10 nm, $\Psi(\lambda)$ and $\Delta(\lambda)$ data were collected at multiple angles of incidence (5 s each at 55$^\circ$, 60$^\circ$, 65$^\circ$) and globally fit to the optical layer model over the wavelength range of $\lambda = 400$--1000 nm.  The layer model consisted of a standard Cauchy layer, $n(\lambda) = A + \frac{B}{\lambda^2} + \frac{C}{\lambda^4}$, for the polymer layer, supported atop a silicon substrate with a native oxide layer.  The 1.4 nm thickness of the native oxide layer was determined from ten measurements across three different 4 inch wafers.  For such thin films, we held the optical constants of the polymer film $A$, $B$, and $C$ fixed at the bulk value for polystyrene (PS) and fit only the film thickness $h$ [1].  Each sample was measured three times at different locations across the 25 mm $\times$ 25 mm film surface with the average of these values was taken to be $h_\text{brush}$.  The grafting density was calculated assuming ideal chain conformations from $\sigma =\frac{\rho \, N_A \, h_\text{brush}}{M_n}$, as described in the main text. 

Pyrene-labeled PS with \Mw = 672 kg/mol and \MwMn = 1.3 was produced by copolymerizing 1-pyrenyl butylmethacrylate at trace levels with styrene, resulting in a fluorescent label content of 1.4 mol\% pyrene.\cite{HuangACSMacroLett2018, BaglayJCP2015}  Fluorescent probe layers of this pyrene-labeled PS with a thickness of $12 \pm 1$ nm were spin-coated from toluene onto freshly cleaved mica and then annealed for $\sim$12 h under vacuum at 120 \degC.  In order to keep the pyrene-labeled PS chains in this probe layer localized during the fluorescence measurements at elevated temperatures ($\leq$170 \degC), the probe layers were lightly crosslinked by exposing them to 254 nm UV light, held at a distance of 16 mm from the film, for 10 min at room temperature, following the protocol used previously in Ref.~\citenum{HuangACSMacroLett2018}.  These 12-nm thick fluorescent probe layers were then floated atop the grafted substrates using a water transfer process [2].  Pieces of the film were also floated onto silicon wafers for film thickness determination by ellipsometry.  To ensure good interpenetration of the pyrene-labeled chains with the end-grafted chains, the samples were then annealed at 170 \degC for 2 hours under vacuum.  

Finally, to isolate the fluorescence signal from interface perturbations due to the free surface, the probe layer was capped with a bulk layer of neat PS.  These caps were made by spin-coating $580 \pm 10$~nm layers of high molecular weight PS with \Mw = 1,920 kg/mol and \MwMn = 1.26 onto mica, separately annealed for $\sim$12 h under vacuum at 120 \degC, before being floated onto the intermixed pyrene-labeled PS / end-grafted PS substrates.  The final annealing step of 20 min at 130~\degC to consolidate the layers and remove any air gaps was done on the fluorometer temperature stage (Instec HCS402) immediately prior to the fluorescence measurements.

\subsection*{Fluorescence measurements and control tests}

Fluorescence measurements, using a Photon Technology International QuantaMaster spectroﬂuorometer, were initiated by stabilizing the samples at 170 \degC for 8--10 min.  Fluorescence intensity at 379 nm, using a 6 nm emission bandpass, was then collected on cooling at 1 K/min over a 3~s window every 30~s, as the pyrene dye was excited at a wavelength of 332 nm with 4 nm excitation bandpass.  This 379 nm emission wavelength\cite{RauscherMacro2013} corresponds to the first peak of pyrene's emission spectrum that is the most sensitive to its local environment [3].  Using plots of the fluorescence intensity $I$ as a function of temperature $T$, linear fits to the $I(T)$ data were performed in the liquid ($T>$ \Tg) and glassy ($T<$ \Tg) regimes, minimizing the mean squared error per degree of freedom within the fitting window, where the glass transition temperature \Tg was determined from the intersection of these linear fits.  

Several control measurements were performed to confirm that the large increase in local \TgZero was only observed when the PS-COOH and PS-OH chains were grafted to the silica substrates.  For example, when films of PS-COOH or PS-OH chains are spin-coated onto the piranha cleaned silica substrates, but not annealed at 170~\degC for 30 min to cause chemical grafting, the PS-COOH and PS-OH chains are readily washed away by the 90~\degC toluene for 20 min bathing conditions we use to remove ungrafted chains, leaving behind a residual $h_\text{brush}$ thickness experimentally equivalent to zero thickness [1].  In addition, when piranha cleaned silica substrates skip the PS-COOH or PS-OH grafting step and the 12-nm-thick pyrene-labeled PS probe layer is floated directly onto the bare silica substrate, we always measure a \TgZero $= 100 \pm 2$ \degC, equivalent to \Tgbulk for PS.  This occurs even when the 12-nm-thick probe layer was annealed on the silica substrate at 170 \degC for 2~h (annealing conditions used to interpenetrate the grafted and homopolymer chains) prior to adding the 600-nm-thick neat PS layer and making the fluorescence measurement as usual.  Such \TgZero control measurements on bare silica are included in Fig.~\ref{Fig3} as gray squares at zero grafting density ($\sigma = 0$).  We also confirmed that UV crosslinking of the probe layer did not alter the measured \TgZero values.\cite{HuangACSMacroLett2018} 

\begin{figure}[b!]
\centering
\includegraphics[width=.6\linewidth]{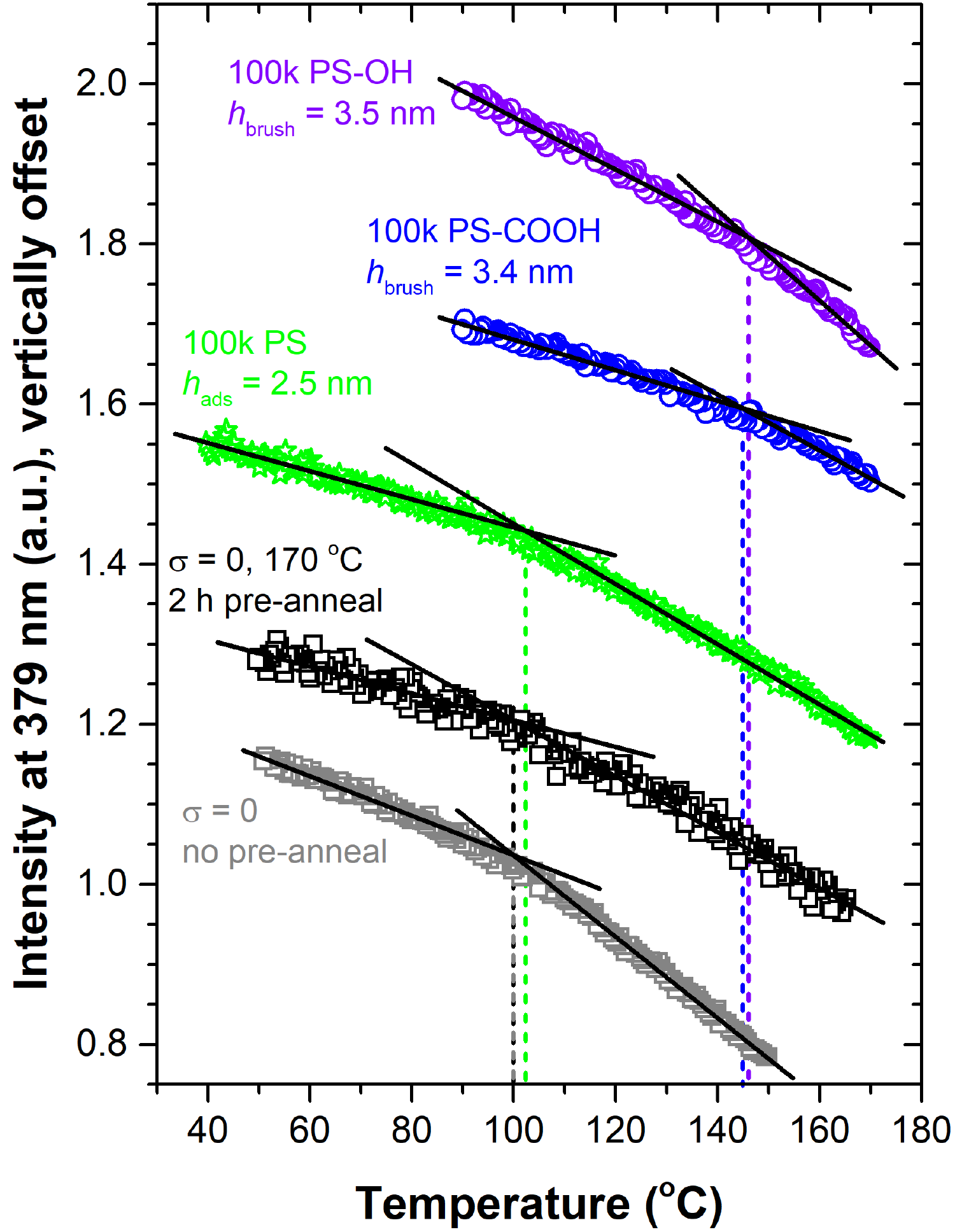}
\captionsetup{width=.8\linewidth}
\caption{\normalsize Normalized fluorescence intensity vs.\ temperature for various sample types, vertically shifted for clarity:  An adsorbed layer formed with 100k neat PS ($h_\text{ads} = 2.5$~nm, \TgZero = 102 \degC, green stars) is compared to 100k PS-COOH ($h_\text{brush} = 3.4$~nm, \TgZero = 145 \degC, blue circles) and 100k PS-OH ($h_\text{brush} = 3.5$~nm, \TgZero = 146 \degC, purple circles), all made with identical sample preparation steps.  Data representing control tests where the pyrene-labeled probe layer was floated directly onto bare, piranha cleaned quartz (corresponding to $\sigma = 0$) are shown for a sample either annealed at 170 \degC for 2 hours under vacuum (black squares) or a sample without this pre-annealing step (gray squares), prior to the addition of the bulk PS cap and fluorescence measurement, resulting in \TgZero values equivalent to \Tgbulk = $100 \pm 2$ \degC.  Thus, only the samples where chemical grafting occurred exhibit a significantly elevated \TgZero relative to that of bulk PS.}
\label{FigS1}
\end{figure}

To verify that the \TgZero increase is caused by chemical grafting and not some physical adsorption process, we also performed control tests where we took 100k neat PS (\Mn = 97.7 kg/mol, \MwMn = 1.04), without the -COOH or -OH grafting groups, and made samples following all the same preparation steps that are used to create grafted layers.  After annealing at 170 \degC for 30 min and washing off the unattached chains by bathing in 90 \degC toluene for 20 min, the equivalent annealing and washing conditions used for making grafted layers, we were left with a residual adsorbed layer $h_\text{ads} = 2.5 \pm 0.5$~nm.  The same sample preparation steps used for the grafted layers were continued by adding the pyrene-labeled layer and annealing at 170 \degC for 2~h to interpenetrate the two layers, followed by capping the film with a bulk $\approx$600~nm layer.  The local \TgZero measured next to such adsorbed layers were found to be equivalent to \Tgbulk, giving \TgZero $= 102 \pm 2$ \degC.  Such an adsorbed layer thickness $h_\text{ads} = 2.5$ nm is comparable to our grafted brush layer thicknesses, $h_\text{brush}$ varied from 0.6 nm to 6.6 nm for the 100k PS-COOH and PS-OH chains corresponding to grafting densities $\sigma$ = 0.004 to 0.042 chains/nm$^2$, but such noncovalent adsorption does not result in an increase in local \TgZero relative to bulk.  In addition, as these adsorbed layer samples underwent all the same sample processing steps as for grafted samples, the \TgZero increases cannot be attributed to some other sample processing condition such as a substrate cleaning or solvent washing step. 

\textbf{Figure~\ref{FigS1}} compares temperature-dependent fluorescence intensity curves for several of these different control tests.  Data are shown for 100k PS-COOH, 100k PS-OH, and 100k neat PS, all following the same sample preparation protocol.  The 100k PS-COOH and PS-OH chains lead to grafted layers with $h_\text{brush}$ = 3.4 and 3.5 nm, corresponding to grafting densities of $\sigma$ = 0.022 and 0.023 chains/nm$^2$, resulting in strongly elevated local \TgZero = $145 \pm 3$ \degC and $146 \pm 3$ \degC, as was demonstrated in Fig.~\ref{Fig1}.  In contrast, the 100k neat PS produces a residual adsorbed layer $h_\text{ads} = 2.5$~nm under the same conditions, resulting in a local \TgZero of $102 \pm 2$ \degC.  Also included are two datasets corresponding to bare silica substrates ($\sigma = 0$).  The pyrene-labelled layer was floated onto the bare piranha-cleaned quartz substrate and either annealed at 170 \degC for 2~hours under vacuum to mimic the annealing conditions used to interpenetrate the pyrene-labelled chains with the grafted chains, or without this pre-annealing step, where both such measurements report a \TgZero = $100 \pm 2$ \degC, equivalent to \Tgbulk for PS.  

Thus, we conclude that the large $\approx$45~K increase in local \Tg observed when the PS-COOH and PS-OH chains are grafted to the silica substrate are specifically the result of the covalent bonds formed.

\subsection*{Interpenetration of matrix chains with end-grafted chains}

An important consideration is that varying the molecular weight of grafted chains can also change the ability of the matrix chains to wet and interpenetrate the grafted surface.  There are two factors here that impact the amount the grafted chains interpenetrate with the matrix chains: wetting and kinetics.  Although the transition between the wet and dry brush regimes is somewhat arbitrary, Matsen and Gardiner have defined it based on the homopolymer's ability to penetrate all the way through the brush to the substrate interface [4].  The wet brush regime was defined in their self-consistent field theory calculations as when the dimensionless grafting density $\frac{\sigma N^{1/2}}{b \, \rho_0} \lesssim 0.8$, primarily independent of the ratio of homopolymer to grafted chain molecular weights $\alpha = \frac{M_\text{homo}}{M_\text{graft}}$ for $\alpha \gtrsim 1$.  The grafted chain length $N$ and statistical segment length $b$ are the same as given in the main text, while $\rho_0^{-1} = \frac{\rho}{m_0} = 0.165$~nm$^3$ is the monomer volume.  For the range of grafted chain molecular weights we investigated (Table~\ref{Table1}), $\alpha$ varies between 2.4 and 60, with the pyrene-labeled PS homopolymer chains (\Mn = 517 kg/mol).  The dimensionless grafting density of our samples has one as high as 0.73, corresponding to the highest grafting density obtained for the 9k PS-OH chains, with all others being less than 0.5.  Thus, our samples can be considered to be in the wet brush regime.  With the ``grafting to'' method we are employing, it is not possible to increase the grafting density further.  

\vspace{1mm}
\begin{figure}[b!]
\centering
\includegraphics[width=.55\linewidth]{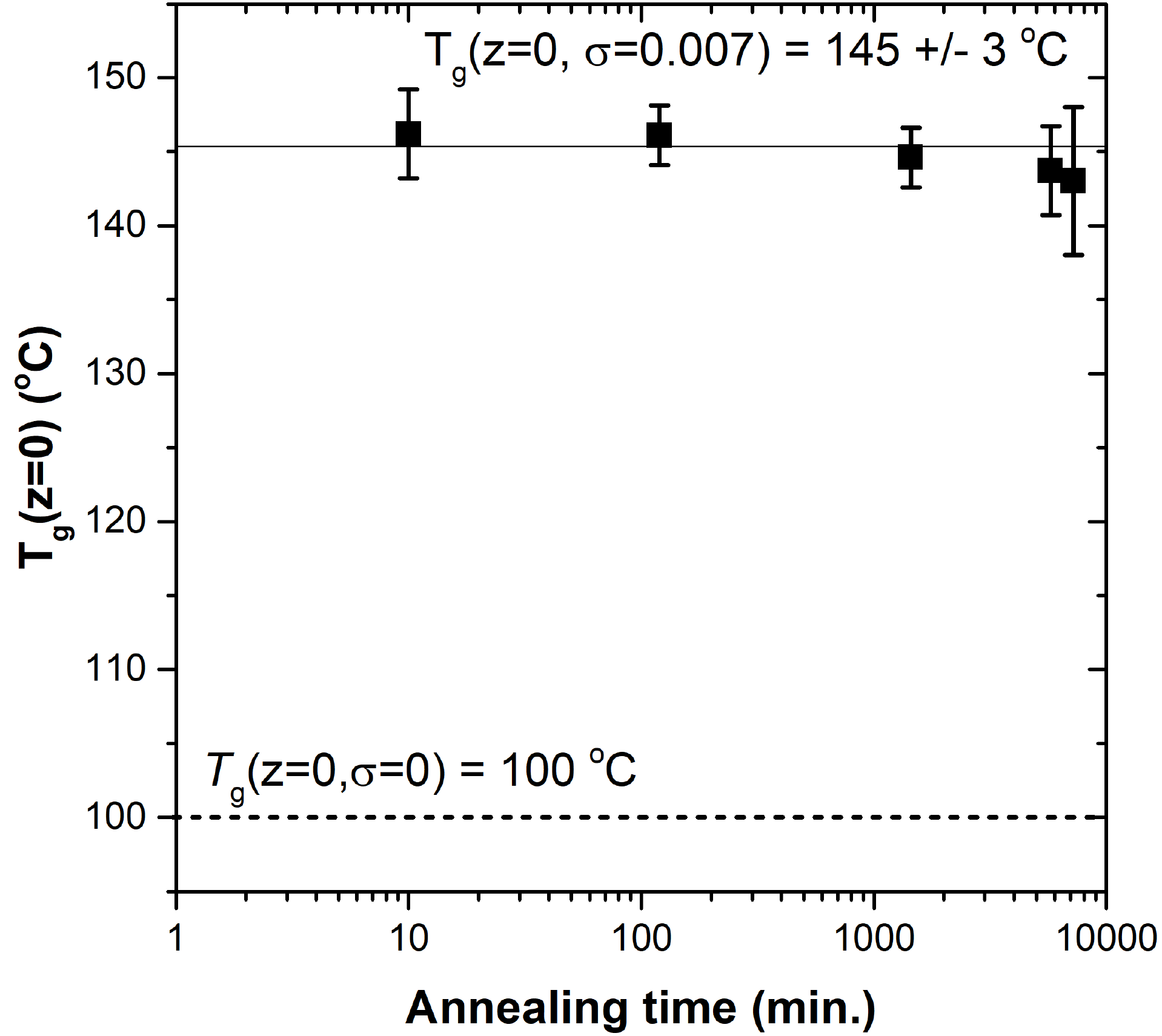}
\captionsetup{width=.8\linewidth}
\caption{\normalsize Measured \TgZero values for 200k PS-COOH grafted chains at $\sigma = 0.007$~chains/nm$^2$ as a function of annealing interpenetration time at 170 \degC, all equivalently elevated at \TgZero = $145 \pm 3$ \degC.  Error bars represent standard deviations of repeated measurements of nominally identical samples. }
\label{FigS2}
\end{figure}

The second factor to consider is the annealing conditions used to weld the homopolymer chains with the grafted chains.  The kinetics of end-grafted chain interpenetration with an overlaying high molecular weight polymer layer can be quite slow,\cite{ChenneviereMacro2013} as the relaxation mechanism of end-grafted chains is limited by the chain-end retraction ``breathing'' modes,\cite{OConnorMcLeishMacro1993} a time scale which grows exponentially with chain length, $t \sim e^N$ [5,6].  This would suggest that samples of the higher molecular weight end-grafted chains such as the 200k PS-COOH could take significantly longer to interpenetrate with the pyrene-labeled homopolymer chains.  However, these homopolymer chains can themselves diffuse by reptation and interpenetrate with the end-grafted chains.  As the reptation time for these chains is only $\sim$5 s at 170 \degC [7], this strongly suggests that our 2~h anneal at this temperature is more than sufficient to ensure the pyrene-labeled homopolymer chains are fully interpenetrated with the grafted chains.  This is consistent with the neutron reflectivity measurements of Chennevière et al.\cite{ChenneviereMacro2013} who investigated the kinetics of interdigitation between h-PS brushes and d-PS melts.  Although they did demonstrate that higher molecular weight end-grafted chains took a considerable length of time to interdigitate with high molecular weight matrices, our 2~h anneal at 170 \degC is an order of magnitude longer than their longest annealing time needed to fully interdigitate 250 kg/mol end-grafted chains with 525 kg/mol matrices.  We have also varied the annealing time at 170 \degC for this interpenetration step between 10 min and up to 120 h (5 days) for the 200k PS-COOH samples with a grafting density of $\sigma = 0.007$~chains/nm$^2$, observing that the measured \TgZero was always elevated at $145 \pm 3$~\degC, a finding that is consistent with the above reasoning that the pyrene-labeled homopolymer chains are fully interpenetrated with the end-grafted chains.  \textbf{Figure~\ref{FigS2}} shows these data, graphing \TgZero as a function of annealing interpenetration time at 170 \degC.  We note that the extremely long annealing times ($>$4 days) at 170 \degC, even though it occurs under vacuum, significantly reduces the fluorescence intensity produced by the sample, indicating that some thermal oxidative degradation of the dyes occur during this process.  Thus, the fluorescence intensity from samples annealed for such extremely long times was substantially noisier, resulting in larger error bars for these measured \Tg values.  

\small 
\vspace{3mm}
\noindent \textbf{Additional References}*  
\hspace{4mm} (*Superscripted references correspond to those listed in the main text.)

\begin{enumerate}[label={[\arabic*]},nosep]
    
    \item  Thees, M. F.; McGuire, J. A.; Roth, C. B.  Review and Reproducibility of Forming Adsorbed Layers from Solvent Washing of Melt Annealed Films. \textit{Soft Matter} \textbf{2020}, \textit{16}, 5366–5387.  \url{https://doi.org/10.1039/d0sm00565g}  
    
    \item  Khodaparast, S.; Boulogne, F.; Poulard, C.; Stone, H. A. Water-Based Peeling of Thin Hydrophobic Films. \textit{Physical Review Letters} \textbf{2017}, \textit{119}, 154502. \url{https://doi.org/10.1103/physrevlett.119.154502}
    
    \item  Valeur, B. \textit{Molecular Fluorescence: Principles and Applications}; Wiley-VCH: Weinheim, 2002.
    
    \item  Matsen, M. W.; Gardiner, J. M. Autophobic Dewetting of Homopolymer on a Brush and Entropic Attraction between Opposing Brushes in a Homopolymer Matrix. \textit{Journal of Chemical Physics} \textbf{2001}, \textit{115}, 2794–2804. \url{https://doi.org/10.1063/1.1385557}
    
    \item  Geoghegan, M.; Clarke, C. J.; Boué, F.; Menelle, A.; Russ, T.; Bucknall, D. G. The Kinetics of Penetration of Grafted Polymers into a Network. \textit{Macromolecules} \textbf{1999}, \textit{32}, 5106–5114. \url{https://doi.org/10.1021/ma982020f}
    
    \item  O’Connor, K.; McLeish, T. Entangled Dynamics of Healing End-Grafted Chains at a Solid/Polymer Interface. \textit{Faraday Discussions} \textbf{1994}, \textit{98}, 67–78. \url{https://doi.org/10.1039/fd9949800067}
    
    \item  Roth, C. B.  \textit{Mobility on Different Length Scales in Thin Polymer Films}; Ph.D. Dissertation, University of Guelph, 2004.
  
\end{enumerate}

\end{document}